\documentclass[prd,a4paper]{revtex4}
\usepackage{graphicx}
\usepackage{dcolumn}
\usepackage{bm}
\usepackage{epsfig}
\usepackage{slashed}
\newcommand{\be}{\begin{equation}}
\newcommand{\ee}{\end{equation}}
\newcommand{\bea}{\begin{eqnarray}}
\newcommand{\eea}{\end{eqnarray}}
\newcommand{\bean}{\begin{eqnarray*}}
\newcommand{\eean}{\end{eqnarray*}}

\newcommand{\gapproxeq}{\lower
.7ex\hbox{$\;\stackrel{\textstyle >}{\sim}\;$}}
\newcommand{\lapproxeq}{\lower
.7ex\hbox{$\;\stackrel{\textstyle <}{\sim}\;$}}

\begin{document}

\bibliographystyle{unsrt}

\title{ Possible contributions to $e^+e^- \to J/\psi + \eta_c$ due to intermediate meson rescatterings}

\author{Yuan-Jiang Zhang$^{1,3}$,  Qiang Zhao$^{1,2,4}$}

\affiliation{1) Institute of High Energy Physics, Chinese Academy
of Sciences, Beijing 100049, P.R. China}

\affiliation{2) Department of Physics, University of Surrey,
Guildford, GU2 7XH, United Kingdom}

\author{Cong-Feng Qiao$^{3,4}$}

\affiliation{3) Graduate University of Chinese Academy of
Sciences, Beijing, 100049, China}

\affiliation{4) Theoretical Physics Center for Science Facilities,
CAS, Beijing 100049, China}

\date{\today}

\begin{abstract}

Inspired by the obvious discrepancies between experiment and
non-relativistic QCD (NRQCD) studies of $e^+ e^-\to J/\psi+\eta_c$
at $\sqrt{s}\simeq 10.6$ GeV, we investigate contributions from
intermediate meson loops as long-range interaction transitions to
this process. The intermediate meson loops include $D\bar D(\bar
D^\ast)$, $D\bar D^\ast(D~{\mbox or}~D^\ast)$, $D^\ast \bar
D^\ast(D)$ and corresponding $D_s$ intermediate mesons. With the
constraints from experimental data on the vertex couplings, we find
that the intermediate meson loops account for $2.7\sim 6.7 \ fb$ of
the cross sections within a reasonable range of cut-off energies of
the factor parameter. We also investigate contributions from the
absorptive part and find that it accounts for approximately
$0.58\sim 1.38 \ fb$. These results imply that contributions from
long-range interaction transitions may still play a role in such an
energy region.

\end{abstract}

\maketitle


\vspace{1cm}

\section{Introduction}

Recently, one of the most hottest topics is the significant
discrepancies between experimental and theoretical results for
exclusive double-charmonium production $e^+e^-\to J/\psi + \eta_c$
at the center mass energy of $10.58$ GeV. In 2002, Belle
Collaboration firstly reported the exclusive cross section for
$\sigma [e^+e^- \to J/\psi + \eta_c]\times {\cal B}(\eta_c\to
\geq4\,\, {\mbox {charged}})=33^{+7}_{-6} \pm 9 \ fb$
~\cite{Abe:2002rb}. In 2004, Belle Collaboration updated their
results $\sigma [e^+e^- \to J/\psi + \eta_c]\times {\cal
B}(\eta_c\to > 2\,\, {\mbox {charged}})=25.6\pm 2.8 \pm 3.4 \ fb$
~\cite{Abe:2004ww} and  BABAR Collaboration also measured the same
quantity, and found $\sigma [e^+e^- \to J/\psi + \eta_c]\times {\cal
B}(\eta_c\to > 2\,\, {\mbox {charged}})=17.6\pm 2.8 \pm 2.1 \
fb$~\cite{Aubert:2005tj}.

Theoretically, based on nonrelativistic QCD(NRQCD) factorization
approach~\cite{Bodwin:1994jh} at leading order (LO) in the QCD
coupling constant ($\alpha_s$) and charm-quark relative velocity
($v$), the predictions given by Braaten and
Lee~\cite{Braaten:2002fi}, Liu, He and Chao~\cite{Liu:2002wq}, and
Hagiwara, Kou and Qiao \cite{Hagiwara:2003cw} are about $2.3\sim 5.3
\ fb$, which are about one order of magnitude  smaller than the
experimental measurements. In order to solve this puzzle, many
solutions have been proposed. In Ref.~\cite{Zhang:2005cha}, the
authors consider the corrections of next-to-leading order (NLO) in
$\alpha_s$, which enhanced the cross section with a $K$ factor (the
ratio of LO plus NLO to LO) of about $1.8\sim 2.1$. Authors of
Ref.~\cite{Braaten:2002fi} find that including relativistic
corrections (order of $v^2$) will lead to the $K\sim
2.0^{+10.9}_{-1.1}$, but with large uncertainties already bared with
the NRQCD matrix elements. In order to reduce the uncertainties of
the order-$v^2$ matrix elements, a potential model is applied to
calculate the quarkonium wave function~\cite{Bodwin:2006dn}, by
which the relativistic correlations and corrections of NLO in
$\alpha_s$ can be treated~\cite{Bodwin:2006ke}, and a cross section
of $17.5\pm 5.7 \ fb$ is obtained. In Ref.~\cite{Bodwin:2007ga}, the
authors present their results by resummation of relativistic
corrections which contains several refinements, such as
nonperturbative NRQCD matrix elements, inclusion of the effects of
the running of $\alpha_s$ etc.  They conclude that the discrepancies
between the theoretical and experimental can be understood. By
determining the matrix elements in a different way,
Ref.~\cite{He:2007te} obtains a value of $20.04 \ fb$. A complete
computation of leading and NLO contributions are recently presented
by Gong and Wang~\cite{Gong:2007db}, which is a directly
confirmation of results from Ref.~\cite{Zhang:2005cha}. Treatments
using light-front approach such
as~\cite{Bondar:2004sv,Ma:2004qf,Braguta:2005kr,Braguta:2006nf} also
provide other possible solutions.

In brief, the present theoretical studies show that NLO
contributions turn out to be important in this double-charmonium
production, which simultaneously raises concerns about the
perturbation expansion. On the other hand, large contributions
from NLO corrections suggest that nonperturbative mechanisms may
start to play a role. This corresponds to the long-range part of
the strong interactions, among which intermediate meson loops
could be a natural explanation for the cross section enhancement
in $e^+e^-\to J/\psi +\eta_c$ at $\sqrt{s}\simeq 10.6$ GeV.

Intermediate meson loop (IML), or intermediate meson rescattering
(or sometimes it is presented as final state interaction (FSI) with
on-shell approximation), as one of the important non-perturbative
transition has been investigated extensively in heavy meson decays.
Cheng {\it et al.} studied the long-distance rescattering effects in
hadronic B decays~\cite{Cheng:2004ru}. From the data accumulated at
B factories and CLEO, it was found that soft final-state
rescattering effects played an essential role in B physics. In
Ref.~\cite{Liu:2006dq}, Liu {\it et al.} also found that
contributions from the hadronic loops turned to be important in
charmonium hadronic decays. Recently, a systematic investigation of
the IML effects in quarkonium hadronic decays involving OZI-rule
violations and isospin breaking reveals that IML plays a role in
many places and sometimes can even compete against direct
scatterings~\cite{Li:2007au,Li:2007xr,Li:2008xm}.

In $e^+e^-\to J/\psi +\eta_c$, the absorptive contributions via IML
are supposed to be present, and they do not interfere with the LO
NRQCD transitions. With constraints from experiment on the effective
couplings, we can explore the IML mechanism in $e^+e^-\to J/\psi
+\eta_c$.

The paper is organized as follows: In section II, we describe the
IML model with effective Lagrangians. The numerical results are
given in Section III.  Discussion and summary are given in Section
IV.

\section{The model}
\subsection{Effective Lagrangians}

In reaction $e^+e^- \to J/\psi +\eta_c$, the final $J/\psi$ and
$\eta_c$ can be produced by direct production of two $c\bar{c}$
pairs. This process is taken care of by pQCD transitions. In our
model, we consider indirect production process where $J/\psi$ and
$\eta_c$ are produced by the intermediate meson loops as shown by
Fig.~\ref{fig-1}.

\begin{figure}
\begin{tabular}{cc}
\includegraphics[height=5cm]{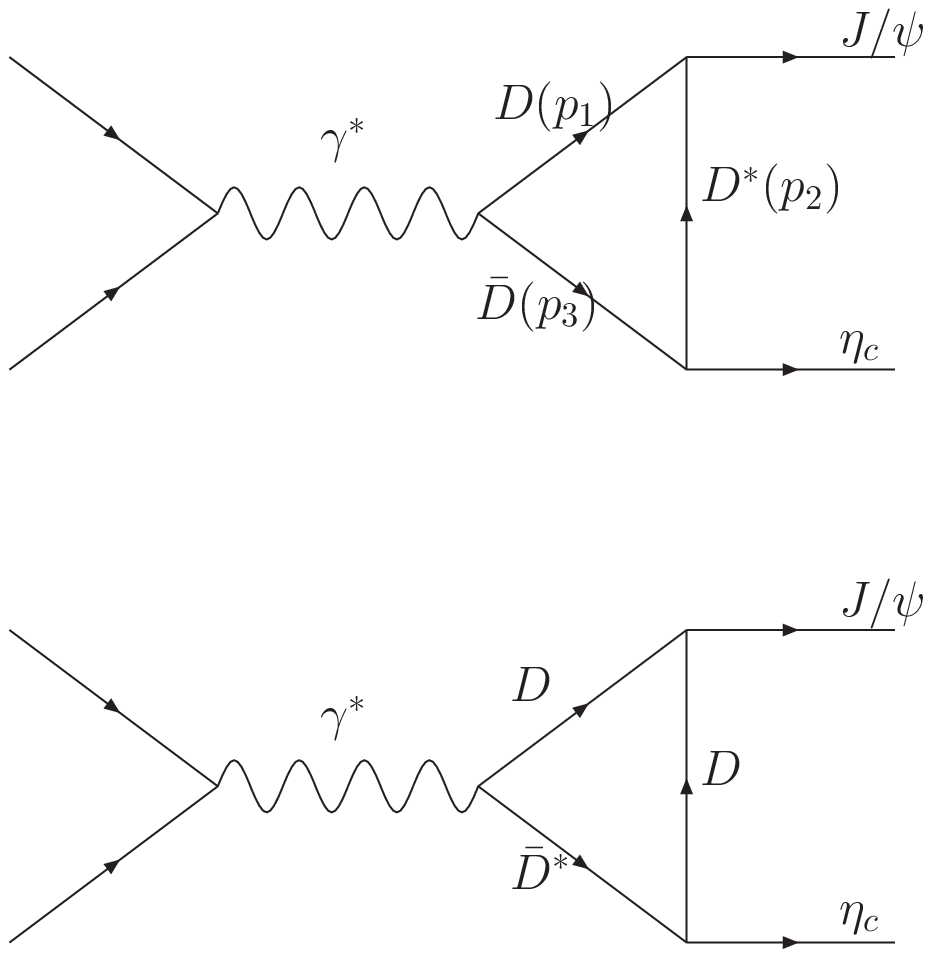}
\includegraphics[height=5cm]{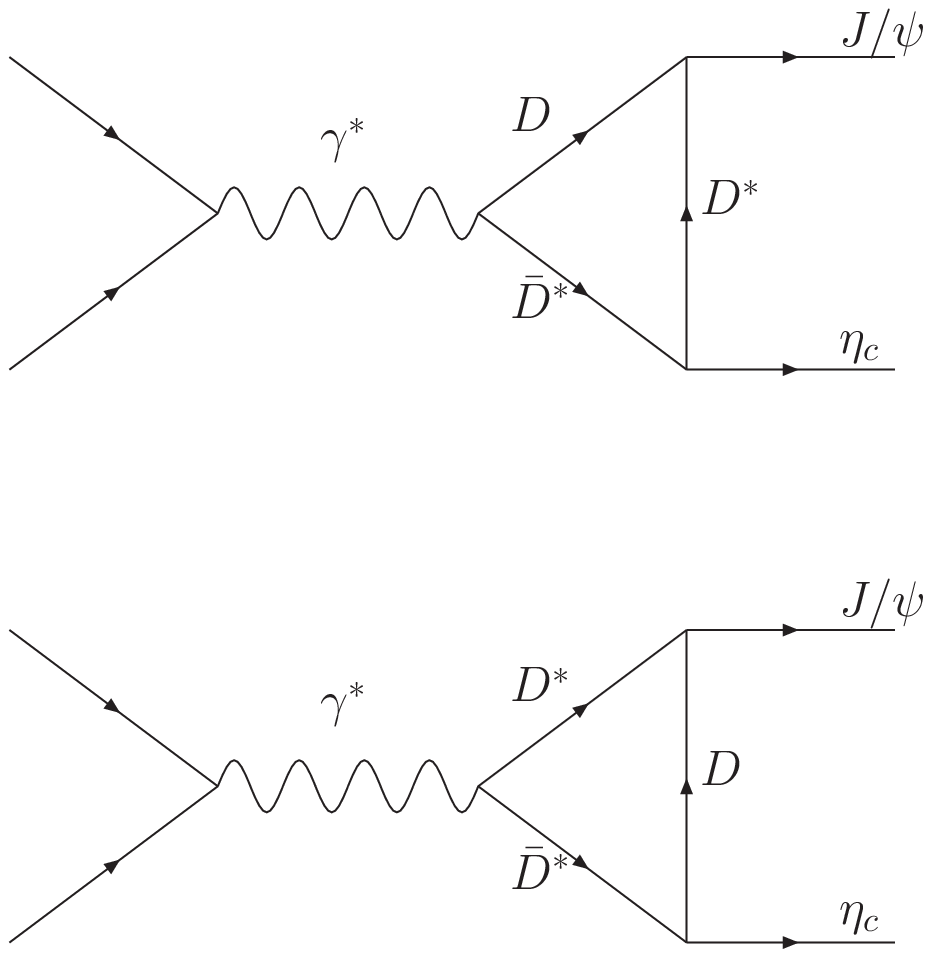}
\end{tabular}
\caption{ Some possible leading intermediate meson loops contribute
to $e^+e^- \to J/\psi + \eta_c$.}\protect\label{fig-2}
\end{figure}

In principle, we should include all the possible intermediate meson
loops in the calculation. But, the break-down of the local
quark-hadron duality allows us to pick up the leading contributions
as a reasonable approximation in the practical
calculation~\cite{Lipkin:1986bi,Lipkin:1986av}. Some possible
leading IML contributions are shown in Fig.~\ref{fig-2}. Since those
intermediate mesons can be off-shell at $\sqrt{s}\simeq 10.6$ GeV,
apart from the absorptive part of the loop transition amplitudes,
the real part can also contribute. This is different from the FSI
approach where only the absorptive amplitudes are considered. In
this work, we shall investigate both.

In order to evaluate the diagrams, we adopt the following
effective Lagrangians:
\bea \nonumber \mathcal{L}_{\gamma D\bar D} &=&
g_{\gamma D\bar D}\{D
\partial_\mu {\bar D}-\partial_\mu D\bar{D} \}\mathcal{A^\mu},\\
\nonumber \mathcal{L}_{\gamma D {\bar D}^\ast} &=& -i g_{\gamma D
{\bar D}^\ast} \epsilon_{\alpha\beta\mu\nu}
\partial^\alpha \mathcal{A^\beta} \partial^\mu \bar {D^\ast}^\nu D + h.c. ,
 \\ \nonumber
\mathcal{L}_{\gamma D^\ast{\bar D}^\ast}&=& g_{\gamma D^\ast{\bar
D}^\ast}\{ \mathcal{A}^\mu(\partial_\mu {D^\ast}^\nu
\bar{D^\ast}_{\nu}-{D^\ast}^\nu\partial_\mu {\bar {D^\ast}_{\nu}})
\\\nonumber &+& (\partial_\mu \mathcal{A}_\nu
{D^\ast}^\nu-\mathcal{A}_\nu\partial_\mu {D^\ast}^\nu)\bar
{D^\ast}^\mu +
{D^\ast}^\mu(\mathcal{A}^{\nu}\partial_\mu\bar{D^\ast}_\nu-\partial_\mu
\mathcal{A}_\nu \bar{D^\ast}^\nu)\},
\\ \nonumber \mathcal{L}_{\psi
D\bar D} &=& g_{\psi D\bar D}\{D \partial_\mu {\bar
D}-\partial_\mu
D\bar{D} \}\mathcal{\psi^\mu}, \\
\mathcal{L}_{\psi D\bar D^\ast} &=& -i g_{\psi D \bar D^\ast}
\epsilon_{\alpha\beta\mu\nu}
\partial^\alpha \mathcal{\psi ^\beta} \partial^\mu \bar{D^\ast}^\nu  D + h.c. .
\label{effective-lagrangians}
\eea
where $\mathcal{A}^\mu$ and $\psi^\mu$ are photon and $J/\psi$
vector-meson fields, respectively, and $D$ is the
pseudoscalar-meson field with $D =(D^0,D^+,D_s^+)$ and $\bar D =
(\bar{D}^0,D^-,D_s^-)^T $; $D^\ast $ is the vector-meson field
with $D^\ast =({D^\ast}^0,{D^\ast}^+,{D_s^\ast}^+)$ and $\bar D =
(\bar{ D^\ast}^0,{D^\ast}^-,{D_s^\ast}^-)^T $. The coupling
constants appearing in the above equations will be determined as
follows.

For the couplings of photon with $D$-meson, such as $g_{\gamma D\bar
D}$, $g_{\gamma D \bar D^\ast}$, $g_{\gamma D^\ast \bar D^\ast}$, we
can extracted them from the experimental data~\cite{Abe:2003et}.
Applying the effective Lagrangians, the $\gamma$ couplings can be
obtained: \bea \nonumber g_{\gamma D\bar D}&=& [\frac
{3s^{5/2}\sigma_{e^+e^-\rightarrow D\bar D}}{2\alpha_e
|\vec{P}_D|^3}]^{1/2},
\\\nonumber
g_{\gamma D\bar D^\ast}&=&[\frac
{16s^{5/2}\sigma_{e^+e^-\rightarrow
D{\bar D^\ast}}}{{\alpha_e |{\vec P}_{D^\ast}|}C_{D\bar D^\ast}}]^{1/2}, \\
g_{\gamma D^\ast{\bar D}^\ast}&=&[\frac
{16s^{5/2}\sigma_{e^+e^-\rightarrow{D^\ast}{\bar D^\ast}}
m_{D^\ast}^4}{\alpha_e |{\vec P}_{D^\ast}^\prime| C_{{D^\ast}\bar
D^\ast}}]^{1/2}, \eea with \bea \nonumber C_{D\bar
D^\ast}&=&m_{\bar D^\ast}^4-2(m_D^2+s)m_{\bar
D^\ast}^2+(m_D^2-s)^2 +\frac{4}{3}s |\vec{P}_{\bar D^\ast}|^2\\
C_{D^\ast\bar D^\ast}&=&s^3+6s^2 m_{D^\ast}^2-28s
m_{D^\ast}^4-48m_{D^\ast}^6-\frac{4}{3}(12m_{D^\ast}^4-6s
m_{D^\ast}^2+s^2)|\vec{P}_{D^\ast}^\prime|^2, \label{coefficient}
\eea
 where ${\vec P_D}$, ${\vec P_{D^\ast}}$, and ${\vec
P_{D^\ast}^\prime}$ are the final state three-momentum for the
processes $e^+e^-\to D\bar{D}$, $D\bar{D^\ast}$, $D^\ast
\bar{D^\ast}$, respectively; $\alpha_e = 1/137$ is the fine
structure-constant; $\sigma$ are the corresponding cross sections.
In the SU(3) limit, we have the following relation $g_{\gamma
D_s\bar D_s}=g_{\gamma D\bar D}$, $g_{\gamma D_s\bar
D_s^\ast}=g_{\gamma D\bar D^\ast}$ and $g_{\gamma D_s^\ast \bar
D_s^\ast}=g_{\gamma D^\ast \bar D^\ast}$. The values are listes in
Table~\ref{table-1}.

 For the $J/\psi D \bar{D^\ast}$
coupling, we use the relation in the heavy quark mass limit
~\cite{Oh-Liu-Ko,Deandrea:2003pv}, \bea
 g_{\psi D  \bar{D^\ast} }=g_{\psi D \bar{D}}/\tilde{M_D},
\eea where $\tilde{M_D}$ is the mass scale of the $D/ D^\ast$
mesons. We adopt the  coupling constants $ g_{\psi D  \bar{D}} =7.44
$ and $ g_{\psi D  \bar{D^\ast}} =3.84$ GeV$^{-1}$ from
Ref.~\cite{Oh-Liu-Ko}, and $g_{{\eta_c} D^\ast \bar{D^\ast}}
=g_{\psi D \bar{D}}$, $g_{{\eta_c} D \bar{D^\ast} } = g_{\psi D
\bar{D^\ast}}$ in the following calculation.

\subsection{Intermediate meson loops contribution}

We take $D\bar D(\bar D^\ast)$, $D\bar D^\ast(D~{\mbox or}~D^\ast)$,
$D^\ast \bar D^\ast(D)$ and corresponding $D_s$ intermediate mesons
into account. The transition amplitude for $e^+e^- \to J/\psi
+\eta_c$ via intermediate meson loops can be expressed as follows:
\bea\label{transition-matrix} M_{fi}^{Loop}= {\bar
v}^{(s^\prime)}(p_e^\prime)(-ie\gamma_\rho)u^s(p_e) \frac
{i\varepsilon_\gamma^\rho} {s} \int \frac {d^4p_2}{(2\pi)^4}
 \frac {T_1T_2T_3}{a_1a_2a_3}{\cal F}(p_2^2),
 \eea
 where $T_{1,2,3}$ are the vertex
functions of which the detailed expressions have been given in
Refs.~\cite{Li:2007au,Li:2008xm}. We include them in the Appendix
for all the above mentioned loops. Four-vector momentum $p_2$ is
for the exchange meson. Variables $a_1\equiv p_1^2-m_1^2,
a_2\equiv p_2^2-m_2^2$, and $a_3\equiv p_3^2-m_3^2$ are the
denominators of the propagators of the intermediate mesons,
respectively.

In the above equation a form factor ${\cal F}(p_2^2)$ is
introduced to take into account the off-shell effects of the
exchanged mesons and also kill the divergence of the integrals.
The following commonly-used form factor is adopted:
\be
 {\cal F}(p^2) = \left(\frac {\Lambda^2 - m_{ex}^2}{\Lambda^2 - p^2}\right)^n,
\ee where $n=0, 1, 2$ correspond to different treatments of the loop
integrals. $\Lambda$ is the cutoff energy, which should not be far
away form the physical mass of the exchanged particles. Since there
are different particles exchanged in the meson loops, a useful
parameterization is as follows~\cite{Cheng:2004ru}: \bea \Lambda =
m_{ex} + \alpha \Lambda_{QCD}, \eea where $\Lambda_{QCD}=220$ MeV
and $\alpha$ is a tunable parameter; $m_{ex}$ is the mass of the
exchanged meson. This way of parameterizing the cut-off energies is
slightly different from the previous
works~\cite{Li:2007au,Li:2008xm,Li:2007xr}, where $\Lambda$ was
universal for all the meson loops. Here, the mass differences
between the exchanged particles are taken into account. The
parameter $\alpha$ will then be constrained by experimental data. In
this work, we apply the dipole form factor in the calculation, i.e.
$n=2$. The relation between the loop amplitudes with the dipole and
monopole form factors are given in Appendix III.

\subsection{ Absorptive contributions from the IML}

It is important to examine the absorptive part of the IML
transitions. Non-negligible contributions from the absorptive part
would be a strong evidence for possible long-range transitions in
the double-charmonium productions.

Taking $D(p_1)$ (or $D^\ast (p_1)$) and $\bar D(p_3)$  (or $D^\ast
(p_3)$) as on-shell particles, we obtain the absorptive transition:
 \bea M_{fi}^{Abs} \nonumber & = & \frac{1}{2} {\bar
v}^{(s^\prime)}(p_e^\prime)(-ie\gamma_\rho)u^s(p_e) \frac
{i\varepsilon_\gamma^\rho} {s} \\\nonumber  && \times \int \frac
{d^3 \vec{p}_1}{(2\pi)^3 2 E_1} \frac{d^3 \vec{p}_3}{(2\pi)^3 2 E_3}
(2 \pi)^4 \delta^4(p_e^\prime + p_e - p_1 - p_3) \frac
{T_1T_2T_3}{p_2^2 - m_2^2 +i m_2 \Gamma_2}{\cal F}(p_2^2) \\ &=&
{\bar v}^{(s^\prime)}(p_e^\prime)(-ie\gamma_\rho)u^s(p_e) \frac
{i\varepsilon_\gamma^\rho} {s} \int_{-1}^{1} \frac{|\vec{p}_1|d
\text{ cos}\theta}{16 \pi \sqrt{s}}\frac{T_1T_2T_3}{p_2^2 - m_2^2 +i
m_2 \Gamma_2}{\cal F}(p_2^2) ,
 \eea
 where $\theta$ is the angle
between $\vec{p}_1$ and $\vec{p}_\psi$. The kinematic definitions
have been given in Appendix II. The vertex functions $T_{1,2,3}$ and
form factor ${\cal F}(p_2^2)$ are the same as the previous
definitions.

\subsection{EM transition via $\Upsilon(4S)$ intermediate
meson }

The present data from Belle were taken at $\sqrt{s}=10.58$ GeV and
10.52 GeV, respectively. The higher one corresponds to the mass of
$\Upsilon(4S)$ while the lower one is a side-band measurement.
However, the datum samples are not sufficient for determining the
cross section differences between these two energies. Therefore, the
role played by $\Upsilon(4S)$ from the present Belle
results~\cite{Abe:2002rb,Abe:2003et,Abe:2004ww} is still unclear,
and the $\Upsilon(4S)\to J/\psi +\eta_c$ coupling is also unknown.
In Ref.~\cite{Jia:2007hy}, exclusive decays of $\Upsilon(4S)$ to
double-charmonium states were evaluated and the branching ratio for
$\Upsilon(4S)\to J/\psi + \eta_c$ was predicted to be at an order of
$10^{-9}$. At amplitude level, such a small contribution can still
produce some structures at the $\Upsilon(4S)$ mass due to the
destructive interference between the resonance strong decay and
continuum amplitudes.

In our calculation we take into account the contribution of the
$\Upsilon(4S)$. Its interference with the IML amplitudes is useful
for us to examine the model-dependence of the IML transition and
sensitivities of the cross sections to its coupling to
$J/\psi\eta_c$.

Fortunately, the EM coupling for $\Upsilon(4S)\to e^+ e^-$ has been
relatively well measured~\cite{pdg2006}. We can then determine the
$V\gamma^\ast$ coupling $eM_V^2/f_V$ by the vector meson dominance
(VMD) model~\cite{Bauer:1975bw}, \be
\frac{e}{f_V}=\left[\frac{3\Gamma_{V\to e^+ e^-}}{2\alpha_e |{\bf
p}_e|}\right]^{1/2} , \ee where $\Gamma_{V\to e^+ e^-}$ is the
partial decay width, $|{\bf p}_e|$ is the electron three-momentum in
the vector meson rest frame, and $\alpha_e=1/137$ is the
fine-structure constant. It should be noted that this form of
interaction is only an approximation and can have large off-shell
effects arising from either off-shell vector meson or virtual photon
fields. In this approach we consider such effects in the $V\gamma P$
coupling form factor which will then be absorbed into the
energy-dependent widths of the vector mesons.

\begin{figure}
\includegraphics[height=2.5cm]{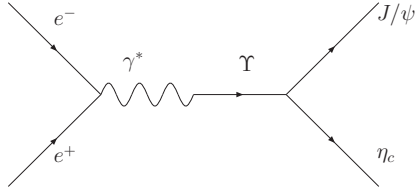}
\caption{The $J/\psi + \eta_c$ production via resonance
$\Upsilon(4S)$.}\protect\label{fig-3}
\end{figure}

The transition amplitude corresponds to Fig.~\ref{fig-3} is
\bea
\nonumber M_{fi} &=&{\bar
v}^{(s^\prime)}(p_e^\prime)(-ie\gamma^\alpha)u^s(p_e) \frac {i}
{s(s-M_\Upsilon^2 + i\Gamma_\Upsilon M_\Upsilon)}\frac
{eM_\Upsilon^2} {f_\Upsilon}g_{\Upsilon\psi\eta_c}
\varepsilon_{\alpha\beta\mu\nu}p_{\eta_c}^\beta
p_\psi^\mu\varepsilon_\psi^\nu \ .
\eea
In case that the partial width of $\Gamma_{\Upsilon\to \psi
\eta_c}$ is available, we can extract the $\Upsilon\psi\eta_c$
coupling via
\bea
\Gamma_{\Upsilon\to \psi \eta_c} = \frac{g_{\Upsilon\psi\eta_c}^2
|{\bf p}_{\psi}|^3 }{12\pi} ,
\eea
where  $|{\bf p}_{\psi}|$ is the three-momentum of $ J/\psi$ in
the $\Upsilon$ meson rest frame.

The total transition amplitudes apart from the leading QCD
contribution can thus be expressed as
\bea
M_{fi}=M_{fi}^{EM}+M_{fi}^{Loop}.
\eea
We estimate the differential cross section via:
\bea
\frac {d\sigma}{d\Omega} = \frac {1} {64\pi^2s} \frac {|P_f|}
{|P_e|} \frac {1}{4} \sum_{spin} |{\cal M}_{fi}|^2,
\eea
where $p_e$ is the three-momentum of the initial electron
(positron) in the overall c.m. system. The mass of the electron
has been neglected, i.e. $p_e=E_{cm}/2$ is applied.

\section{Numerical Results}

As mentioned before, the present experimental data at
$\sqrt{s}=10.52$ and 10.58 GeV cannot tell the role played by
$\Upsilon(4S)$. But we can still define quantity $R$, which is the
$e^+e^- \to J/\psi +\eta_c$ cross section ratio at $\sqrt{s}=10.52$
GeV to that at $\sqrt{s}=10.58$ GeV, i.e. $R \equiv
{\sigma_{10.52}}/{\sigma_{10.58}}$. It is a function of both
resonance direct transition and IML amplitudes, and we expect that
it should not be sensitive to the cut-off energy introduced to the
IML though the exclusive IML contributions may have strong
dependence on it. Unfortunately, no information about the
$\Upsilon(4S)$ coupling to $J/\psi\eta_c$ is available. We then
simply assume $R=0.5$ at $\alpha=1.8$ to fix the coupling
$g_{\Upsilon\psi\eta_c}=5.26\times 10^{-5}$ GeV$^{-1}$. This
corresponds to $BR(\Upsilon\to J/\psi\eta_c)=2.9\times 10^{-7}$. One
should not take this value seriously since our strategy here is to
fix $g_{\Upsilon\psi\eta_c}$, and then examine the sensitivity of
$R$ to the form factor parameter $\alpha$ as a test of the behavior
of the IML.

\begin{figure}
\includegraphics[scale=0.6]{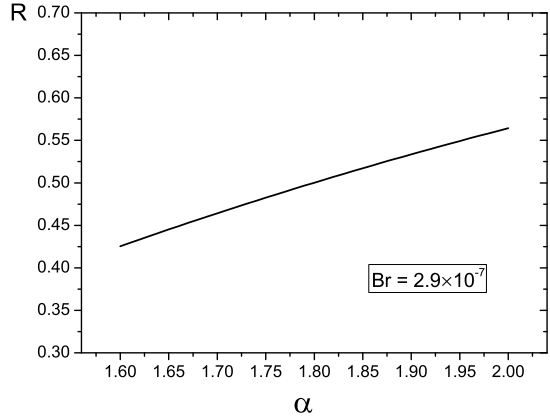}
\caption{ The sensitivity of $R$ to the form factor parameter
$\alpha$ with the $\Upsilon J\psi\eta_c$ coupling fixed at $R=0.5$
and $\alpha=1.8$. }\protect\label{fig-r-alpha}
\end{figure}

In Fig.~\ref{fig-r-alpha}, by fixing
$g_{\Upsilon\psi\eta_c}=5.26\times 10^{-5}$ GeV$^{-1}$, we plot
the ratio $R$ with a varying $\alpha$. It shows that within the
commonly accepted range of $\alpha=1.6\sim 2.0$, the ratio $R$
varies from 0.43 to 0.57. This is an indication of insensitivity
of the loop contributions to the form factors within a reasonable
range of the cut-off energy. We also find that such a property is
retained even for larger $g_{\Upsilon\psi\eta_c}$. In such a case,
the cross section at the mass of $\Upsilon(4S)$ is enhanced by the
resonance contribution while the side-band cross section is
relatively small. A quantitative determination of the
$\Upsilon(4S)$ requires future precise measurements of the cross
sections at both $\sqrt{s}=10.52$ and 10.58 GeV.

\begin{figure}
\includegraphics[scale=0.4]{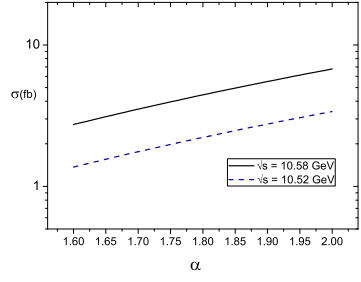}
\caption{ Sensitivities of the $e^+e^- \to J/\psi + \eta_c$ cross
section to the form factor parameter $\alpha$ at $\sqrt{s}=10.58$
(solid line) and $\sqrt{s}=10.52$ GeV (dashed line) with $R=0.5$.
}\protect\label{fig-xsect-both}
\end{figure}

We also plot the cross-section-dependence on $\alpha$ at
$\sqrt{s}=10.52$ and 10.58 GeV in Fig.~\ref{fig-xsect-both}. It
shows that these two cross sections increase with $\alpha$ slowly
and also appear to be stable. Again the difference between these two
cross sections are due to the assumed contribution from
$\Upsilon(4S)$.

\begin{figure}
\includegraphics[scale=0.4]{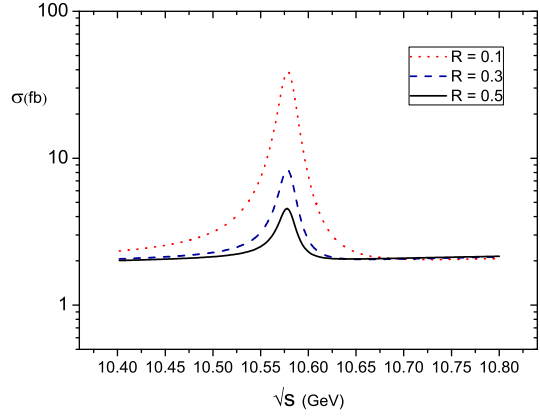}
\caption{ The $\sqrt{s}$ evolution of $e^+e^- \to J/\psi + \eta_c$
cross section with different $R$ values which correspond to
different branch ratios for $\Upsilon(4s)\to \psi+\eta_c$. The form
factor parameter $\alpha = 1.8$ is
adopted.}\protect\label{fig-xsect-r}
\end{figure}

In Fig.~\ref{fig-xsect-r}, we fix the form factor parameter
$\alpha=1.8$ and then investigate the effects from the
$\Upsilon(4S)$ at different $R$ values. The total cross sections
from the IML plus resonance $\Upsilon(4S)$ are presented. A smaller
value of $R$ corresponds to a larger $g_{\Upsilon\psi\eta_c}$
coupling in the present choice of a constructive relative phase. It
is possible that their interference is destructive and a dip will
appear at $\sqrt{s}=10.58$ GeV~\cite{Jia:2007hy}. In this sense, we
emphasize again that it is essential to have precise data for the
cross sections at both $\sqrt{s}=10.52$ and 10.58 GeV.

\begin{figure}
\includegraphics[scale=0.6]{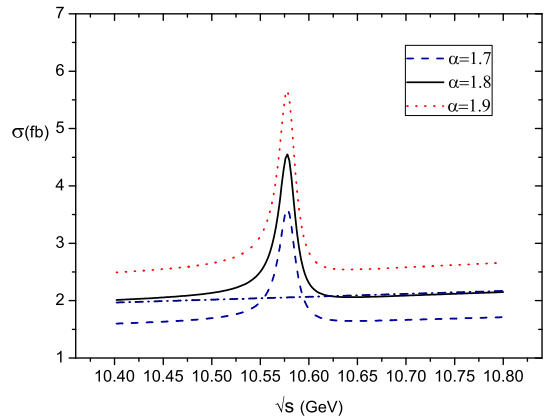}
\caption{ The $\sqrt{s}$ evolution of  $e^+e^- \to J/\psi + \eta_c$
cross section with $R=0.5$ but varying values of $\alpha=1.7$
(dashed line), 1.8 (solid line) and 1.9 (dotted line). The
dot-dashed line denotes the result without $\Upsilon(4S)$
contribution at $\alpha=1.8$. }\protect\label{fig-xsect-alpha}
\end{figure}

In Fig.~\ref{fig-xsect-alpha}, we plot the total cross section
again, but with several different $\alpha$ values. The solid curve
is the same as that in Fig.~\ref{fig-xsect-r} with $\alpha=1.8$.
Again, some sensitivities to $\alpha$ are highlighted. We also
present the results without $\Upsilon(4S)$, i.e. exclusive cross
sections from the IML, as denoted by the dot-dashed curve. A flat
behavior is observed along with $\sqrt{s}$.

It is interesting to see that the total cross section without the
$\Upsilon(4S)$ has the same order of magnitude as the pQCD leading
order results. In Tab.~\ref{tab-2} we list the total cross sections
with different $\alpha$ values. In a range of $\alpha=1.6\sim 2.0$
the total cross sections increase from 2.74 $fb$ to 6.76 $fb$.
Eventually, this range is acceptable for the form factor
uncertainties which is unavoidable in this effective Lagrangian
approach. Similarly, the side-band cross sections at
$\sqrt{s}=10.52$ GeV (i.e. can be viewed as no $\Upsilon(4S)$
contributions) vary from $1.3 \ fb$ to 3.5 $fb$.

\begin{figure}
\includegraphics[scale=0.6]{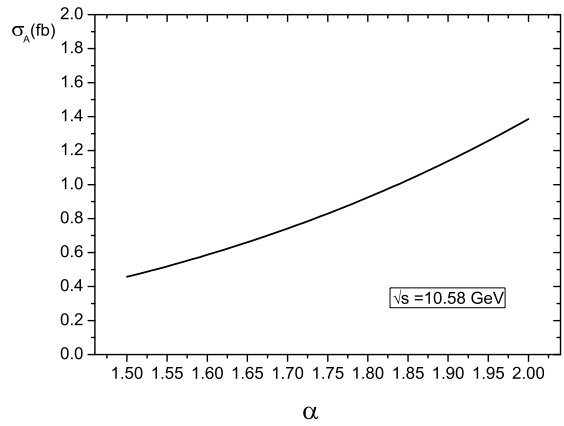}
\caption{The $\alpha$-dependence of the absorptive cross sections
from the meson loops. }\protect\label{fig-abs-alpha}
\end{figure}

In order to further clarify the role played by the IML, we calculate
the absorptive contributions from the loops by making the on-shell
approximation for the intermediate mesons. In
Fig.~\ref{fig-abs-alpha}, the $\alpha$-dependence of the cross
section at $\sqrt{s}=10.58$ GeV is presented. Since the absorptive
part will be present in the transition amplitudes, its stability
within the commonly accepted range of the form factor parameter
turns to be essential for understanding its role in this process. As
shown by Fig.~\ref{fig-abs-alpha}, with $\alpha$ from 1.6 - 2.0, the
absorptive cross section varies from $\sim 0.6 - 1.4 \ fb$, and can
be regarded as quite stable. We also list the cross sections with
different values of $\alpha$ in Tab.~\ref{tab-2} to compare with the
full loop calculations.

\begin{figure}
\includegraphics[scale=0.6]{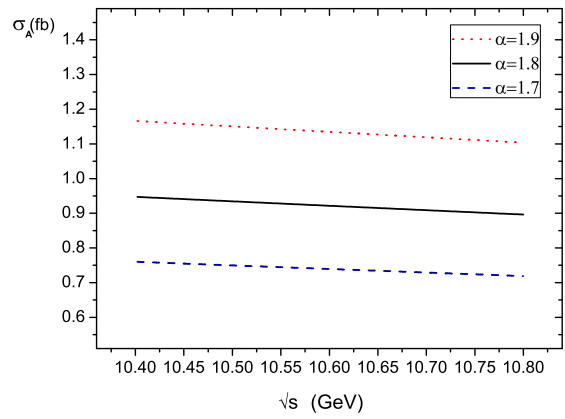}
\caption{The energy evolution of the absorptive cross sections from
the meson loops with different $\alpha$ values.
}\protect\label{fig-abs}
\end{figure}

We then examine the total absorptive cross sections with fixed
$\alpha$ values in Fig.~\ref{fig-abs}. It shows that with
$\alpha=1.8$, the cross section is about 0.95 $fb$, and its
energy-dependence is rather weak. Again, larger values for $\alpha$
produce larger cross sections. The most interesting feature is that
the absorptive cross sections are nearly the same order of magnitude
as the LO results of NRQCD.

One should be cautioned in the understanding of the result of
Fig.~\ref{fig-abs}. It eventually reflects the energy-dependence of
the IML form factors, and with the couplings of $e^+ e^-\to
D\bar{D}$ etc fixed at $\sqrt{s}\simeq 10.56$ GeV. In this approach
QCD hard interactions have been contained in those couplings derived
at $\sqrt{s}\simeq 10.56$ GeV, i.e. $g_{\gamma D\bar{D}}$,
$g_{\gamma D\bar{D^*}}$ and $g_{\gamma D^*\bar{D^*}}$. The form
factor in Eq.~(\ref{transition-matrix}) will then take care of the
soft interactions arising from the intermediate meson loops. Hence,
one should not compare the energy-dependence of Fig.~\ref{fig-abs}
with the QCD factorization~\cite{Bodwin:2008nf,Zhang:2005cha} for
the direct production of double charmonia.

Also, it should be pointed that the IML contributions as a source of
long-range interaction transition are different from the soft
exchanges in the pQCD factorization~\cite{Bodwin:2008nf}. It is not
a full correspondence of the soft QCD. Ideally, we expect that the
sum of all the possible intermediate meson loops be equivalent to
the soft exchanges in the pQCD factorization according to the
quark-hadron duality argument~\cite{Lipkin:1986bi,Lipkin:1986av}. In
reality the break-down of the local quark-hadron duality allows us
to pick up the leading IML contributions as a first-order
approximation.

\section{Summary}
 We estimate the intermediate meson loop contributions to
$e^+e^- \to J/\psi + \eta_c$ in an effective Lagrangian theory. By
applying the available experimental information to the constraints
of the meson-meson coupling vertices, we investigate contributions
from $D\bar{D}(D^*)$, $D\bar{D^*}(D)$, $D\bar{D^*}(D^*)$ and
$D^*\bar{D^*}(D)$. The model-dependence mainly comes from the form
factors adopted for the loop integrals. Fortunately, it shows that
the results do not vary dramatically with the cut-off energies
within the commonly accepted range. The resonance $\Upsilon(4S)$
effects are also estimated. Precise measurements of the cross
sections at the resonance energy and side-band will be able to
provide useful information on the $\Upsilon(4S)$.

In order to clarify the IML contributions, we also calculate the
absorptive part of the loops and find that the IML contributes
nearly the same order of magnitude as the LO of NRQCD. It is likely
that the long-range IML be an important mechanism apart from the
NRQCD LO transitions. Qualitatively, the long-range IML transitions
may have some overlaps with the NLO processes if they are not
obviously suppressed. Note that the IML contributions go to the
effective coupling of the anti-symmetric tensor at hadronic level.
It suggests that the relativistic corrections may also have some
overlaps with the IML mechanism. To understand the large cross
sections from the experiment, one perhaps should consider both short
and long-range transitions to obtain an overall consistent
prescription. Again, we emphasize the importance of extracting the
precise cross sections at and off resonance $\Upsilon(4S)$.

\section*{Acknowledgement}

We would like to thank C.H. Chang, K.T. Chao, T. Huang, Y. Jia, X.Q.
Li, S. Olsen, W. Wang, and J.X. Wang for fruitful discussions. We
also thank G. Li for double-checking part of the calculations and
useful discussions. This work is supported, in part, by the National
Natural Science Foundation of China (Grants No. 10675131), Chinese
Academy of Sciences (KJCX3-SYW-N2), and the U.K. EPSRC (Grant No.
GR/S99433/01).

\section*{Appendix I: Vertex functions for the intermediate meson
loops}

\begin{figure}
\includegraphics[scale=0.6]{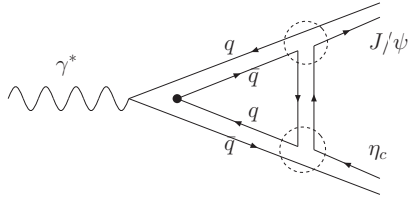}
\caption{ The schematic diagram for the long-range IML contributions
to $J/\psi+\eta_c$ production in $e^+ e^-$
annihilation.}\protect\label{fig-1}
\end{figure}

For  $D\bar{D}(D^\ast)$, the vertex functions are
\be
\left\{\begin{array}{ccl}
 T_1 &\equiv& ig_1(p_1-p_3)\cdot \varepsilon_\gamma \\
 \nonumber
 T_2&\equiv& ig_2\varepsilon_{\alpha\beta\mu\nu}p_{\psi}^\alpha\varepsilon_{\psi}^\beta
 p_2^\mu\varepsilon_2^\nu \\
 T_3&\equiv& ig_3(p_{\eta_c}+p_3)\cdot \varepsilon_2\end{array}\right.
 \ee
where $g_1$, $g_2$, and $g_3$ are the coupling constants at the
meson interaction vertices (see Fig. \ref{fig-1}(a)). The four
vectors, $p_{J/\psi}$, and $p_{\eta_c}$ are the momenta for the
final state $J/\psi$ and $\eta_c$ meson; The four-vector momentum,
$p_1$, $p_2$, and $p_3$ are the intermediate mesons, respectively.

As shown by Fig.~\ref{fig-1}, the vertex functions for the
$D\bar{D^\ast}(D)+c.c.$ loop are \be \left\{
\begin{array}{ccl}
 T_1 &\equiv &i f_1
 \varepsilon_{\alpha\beta\mu\nu}
 p_{3}^\alpha \varepsilon_{3}^\beta p_\gamma^\mu \varepsilon_\gamma^\nu \ , \nonumber \\
 T_2&\equiv& i f_2(p_2-p_1)\cdot \varepsilon_{\psi} \ , \nonumber\\
 T_3&\equiv & i f_3(p_{\eta_c}-p_2)\cdot \varepsilon_3 \ .
 \end{array}\right.
\ee where $f_{1,2,3}$ are the coupling constants.

We also consider the transition amplitude from the intermediate
$D\bar{D^\ast}(D^\ast)+c.c.$ loop , which can be expressed as the
following formula:
 \be\label{amp-loop-3}
\left\{\begin{array}{ccl}
 T_1 &\equiv & i h_1
 \varepsilon_{\alpha\beta\mu\nu}
 p_\gamma^\alpha \varepsilon_\gamma^\beta p_3^\mu \varepsilon_3^\nu \ , \\
 \nonumber
 T_2&\equiv&  i h_2
 \varepsilon_{\alpha^\prime\beta^\prime\mu^\prime\nu^\prime}
 p_2^{\alpha^\prime} \varepsilon_2^{\beta^\prime} p_{\psi}^{\mu^\prime} \varepsilon_{\psi}^{\nu^\prime} \ , \\
 T_3&\equiv &  i h_3
 \varepsilon_{\alpha^{\prime\prime}\beta^{\prime\prime}\mu^{\prime\prime}\nu^{\prime\prime}}
 p_2^{\alpha^{\prime\prime}} \varepsilon_2^{\beta^{\prime\prime}} p_3^{\mu^{\prime\prime}} \varepsilon_3^{\nu^{\prime\prime}}
\end{array}\right.
 \ee
where $h_{1,2,3}$ are the coupling constants.

Transition amplitude from the intermediate
$D^\ast\bar{D^\ast}(D)+c.c.$ loop can be writen as
Eq.~(\ref{amp-loop-4}) \be
\label{amp-loop-4}\left\{\begin{array}{ccl}
 T_1 &\equiv & i \lambda_1
[-\varepsilon_\gamma\cdot(p_1 - p_3)\varepsilon_1\cdot\varepsilon_3 + 2p_1\cdot\varepsilon_3 \varepsilon_1\cdot\varepsilon_\gamma +2\varepsilon_1\cdot p_3\varepsilon_\gamma\cdot\varepsilon_3] \\
 \nonumber
 T_2&\equiv &  i \lambda_2
 \varepsilon_{\alpha\beta\mu\nu}
 p_1^{\alpha} \varepsilon_1^{\beta} p_\psi^{\mu}
 \varepsilon_\psi^{\nu}\\
 \nonumber
 T_3&\equiv &  i \lambda_3
 (p_{\eta_c}-p_2)\cdot\varepsilon_3

\end{array}\right.
 \ee
where $\lambda_{1,2,3}$ are the coupling constants.

\section*{Appendix II: Four-vector momentum expressions and wave
functions}
 In order to calculate the absorptive
amplitude of $e^+e^- \to J/\psi+ \eta_c$, it is convenient if we
choose the overall c.m. frame and choose the $z$ axis along the
three-vector momentum of $J/\psi$. After taking $m_e = m_{e^+}\simeq
0 $, we have
 \be \left\{\begin{array}{ccccccccl} p_{\psi}^\mu
&=& (E_\psi,0,0,|\vec{p}_f|)^T \\
p_{\eta_c}^\mu&=& (E_{\eta_c},0,0,-|\vec{p}_f|)^T \\
p_1^\mu &=& (E_1, |\vec{p}_1|\sin\theta \cos\varphi,
|\vec{p}_1|\sin\theta \sin\varphi,|\vec{p}_1|\cos \theta)^T
\\
p_2^\mu &=& (E_2, -|\vec{p}_1|\sin\theta \cos\varphi,
-|\vec{p}_1|\sin\theta \sin\varphi,|\vec{p}_f|-|\vec{p}_1|\cos
\theta)^T
\\
p_3^\mu &=& (E_3, -|\vec{p}_1|\sin\theta \cos\varphi,
-|\vec{p}_1|\sin\theta \sin\varphi,-|\vec{p}_1|\cos \theta)^T\\
p_e^\mu &=& (E,E\sin\theta_1
\cos\varphi_1,E\sin\theta_1\sin\varphi_1,E\cos
\theta_1)^T\\
{p_e^\prime}^\mu &=& (E,-E\sin\theta_1
\cos\varphi_1,-E\sin\theta_1\sin\varphi_1,-E\cos
\theta_1)^T\end{array}\right. \ee
 and
 \be
\left\{\begin{array}{cccccl} \nonumber \epsilon_\psi^\mu(\pm1)&=&
\mp\frac{1}{\sqrt{2}}(0,1,\pm i,0)^T
\\\nonumber \epsilon_\psi^\mu(0) &=&
(|\vec{p}_f|/M_\psi,0,0,E_\psi/M_\psi)^T\\
u^s(p_e) &=& \frac{\slashed{p}_e}{\sqrt{E}}
(\chi(s)^T,0,0)^T \\
v^s(p^\prime _e) &=& \frac{-\slashed{p}^\prime_e}{\sqrt{E}}
(0,0,\chi(s)^T)^T \end{array}\right. \ee
 where $\chi(s)$ is the wave function of the electron and positron; $(\theta_1,\varphi_1)$
 are the azimuth angles opened by the three-vector momentum.
 \be \chi(+\frac{1}{2}) = \left(
\begin{array}{cccccl} \cos \frac{\theta_1}{2} \\
\sin \frac{\theta_1}{2} e^{i\varphi_1}\end{array}\right),\ \ \ \ \
\ \chi(-\frac{1}{2}) = \left(
\begin{array}{ccccl} -\sin \frac{\theta_1}{2} e^{-i\varphi_1} \\
\cos \frac{\theta_1}{2}\end{array}\right).\ee

\section*{Appendix III: Useful formula }

Started with the case of no form factor in the integral,
Eq.~(\ref{transition-matrix}) can be expressed as the power of the
four-vector momentum of the exchange-meson ($p_2$), and it is easy
to shown that the power of $p_2$ is no more than three. We apply
Feynman parameter dimensional regularization scheme to express the
integral as a linear combination of the following forms:
 \bea
\nonumber\int\frac{d^4p_2}{(2\pi)^4} \frac{p_2^\mu}{[(p_2-p_\psi)^2
- m_1^2][p_2^2-m_2^2][(p_2+p_{\eta}
^2)-m_3^2]}&=&\frac{i}{16\pi ^2}\int dxdy \frac{-P^\mu}{M^2-P^2}, \\
\nonumber \int\frac{d^4p_2}{(2\pi)^4} \frac{p_2^\mu
p_2^\nu}{[(p_2-p_\psi)^2 - m_1^2][p_2^2-m_2^2][(p_2+p_{\eta}
^2)-m_3^2]}&=& \frac{i}{16\pi ^2}[\int dxdy \frac{P^\mu
P^\nu}{M^2-P^2} +
\frac{1}{4}\Gamma(\frac{\epsilon}{2})\delta^{\mu\nu}],\\ \nonumber
\int\frac{d^4p_2}{(2\pi)^4} \frac{p_2^\mu p_2^\nu
p_2^\sigma}{[(p_2-p_\psi)^2 - m_1^2][p_2^2-m_2^2][(p_2+p_{\eta}
^2)-m_3^2]}&=& \frac{i}{16\pi ^2}[\int dxdy \frac{-P^\mu P^\nu
P^\sigma}{M^2-P^2} \\ &-&
\frac{1}{4}\Gamma(\frac{\epsilon}{2})(\delta^{\mu\nu}P^\sigma +
\delta^{\mu\sigma}P^\nu +\delta^{\nu\sigma}P^\mu)],\eea
 where $P =
y p_{\eta}-x p_\psi , M^2 = x (p_\psi ^2-m_1^2) + y(p_\eta ^2
-m_3^2) -(1-x-y)m_2^2 $; $\epsilon$ is a infinitesimal parameter.
The details can be found in Refs.~\cite{Denner:1991kt,Hahn:1998yk}.

To cancel the divergence with a monopole form factor, we just
replace the exchanged-meson mass by the cut-off energy $\Lambda$ to
obtain the transition amplitude since the divergent term is
independent of the intermediate meson masses:
 \bea \nonumber {\cal
M}_{fi}^{mo} (m_1,m_2,m_3,\Lambda) &=& {\cal M}_{fi}(m_1,m_2,m_3) -
{\cal M}_{fi}(m_1,\Lambda,m_3),
 \eea
 where ${\cal M}_{fi}(m_1,m_2,m_3)$ denotes the amplitude of
 Eq.~(\ref{transition-matrix}) with ${\cal F}(p_2^2)=1$.

For the case of a dipole form factor, we use the identity
 \bea
\left(\frac {\Lambda^2 - m^2}{\Lambda^2 - p^2}\right)^2 =
\lim_{\delta\rightarrow 0} \frac {\Lambda^2 - m^2}{\Lambda^2 -
p^2} \frac {(\Lambda+\delta)^2 - m^2}{(\Lambda+\delta)^2 - p^2},
\eea
 to have
 \bea \nonumber {\cal M}_{fi}^{di}
(m_1,m_2,m_3,\Lambda) &=& {\cal M}_{fi}^{mo}(m_1,m_2,m_3,\Lambda) +
\lim _{\delta\rightarrow
0}\frac{m_2^2-\Lambda^2}{2\Lambda\delta}{\cal
M}_{fi}^{mo}(m_1,\Lambda,m_3,\Lambda + \delta) . \eea

\begin{table}[htbp]
\begin{center}
\begin{tabular}{|c|c|c|}
 \hline coupling constants          & value      & Cross section ($pb$) \\ [1ex]
 \hline   $g_{\gamma D\bar D}$           & 4.81  $\times 10^{-3}$ &          0.04         \\      [1ex] \hline
          $g_{\gamma D\bar D^\ast}$       & 2.73  $\times 10^{-3}$ GeV$^{-1} $             &          0.71          \\ [1ex]  \hline
          $g_{\gamma D^\ast \bar D^\ast}$ & 1.10  $\times 10^{-3}$                         &          0.65          \\ [1ex]  \hline
\end{tabular}
\caption{The coupling constants of photon with D-meson determined in
$e^+e^-\to D\bar D$, $D\bar D^\ast$, $D^\ast \bar D^\ast$. The cross
sections are taken from Ref.~\cite{Abe:2003et}.} \label{table-1}
\end{center}
\end{table}

\begin{table}[htbp]
\begin{center}
\begin{tabular}{|c|c|c|c|c|c|}
\hline\ \ \ \  $\alpha \ \ \ \ $ \ \ \ \ & \ \ \ \ 1.6  \ \ \ \ &\ \ \ \ 1.7 \ \ \ \ &\ \ \ \ 1.8\ \ \ \ &\ \ \ \ 1.9 \ \ \ \   &\ \ \ \ 2.0 \ \ \ \   \\
[1ex] \hline $\sigma_{10.58} (fb)$
 &2.74            & 3.52           &4.43  &5.51 & 6.76  \\
[1ex] \hline $\sigma^\prime _{10.58} (fb) $
 &1.28            & 1.64           &2.06  &2.55 & 3.11  \\
[1ex] \hline $\sigma_{A} (fb)$
 & 0.587          & 0.741          &0.924 & 1.14 & 1.38  \\
[1ex] \hline
\end{tabular}
\caption{$e^+e^- \to J/\psi + \eta_c$ cross sections with
different $\alpha$ values at $\sqrt{s}=10.58$ GeV. The ratio
$R=0.5$ is applied. $\sigma$ is the results from the IML with
$\Upsilon(4S)$ contribution($\sigma^\prime$ without $\Upsilon(4S)$
contribution), while $\sigma_{A}$ is from the absorptive
transitions where no $\Upsilon(4S)$ considered.} \label{tab-2}
\end{center}
\end{table}

\end{document}